\documentclass[reprint,amsmath,amssymb,aps,pra,onecolumn]{revtex4-2}

\usepackage{graphicx}
\usepackage{dcolumn}
\usepackage{bm}
\usepackage{hyperref}
\hypersetup{
    colorlinks=true,
    linkcolor=blue,
    filecolor=blue,      
    urlcolor=blue,
    citecolor=blue,
    pdftitle={}
    }
\usepackage{float}
\usepackage{physics}
\usepackage{upgreek}
\usepackage{multirow}
\usepackage{xcolor}
\usepackage[normalem]{ulem}
\usepackage[shortlabels]{enumitem}

\newcommand{\avgPe}{\langle\mathrm{Pe}\rangle}

\begin{document}

\preprint{APS/123-QED}

\title{Enhancement of swimmer diffusion through regular kicks: analytic mapping of a scale independent parameter space}

\author{Arnau Jurado Romero}
\email{arnau.jurado@upc.edu}
\affiliation{Departament de Física, Universitat Politècnica de Catalunya, \\ Barcelona 08034, Spain}

\author{Carles Calero}
\email{carles.calero@ub.edu}
\affiliation{Departament de Física de la Matèria Condensada and Institut de Nanociència i
Nanotecnologia, \\ Universitat de Barcelona, 08028 Barcelona, Spain}

\author{Rossend Rey}
\email{rosendo.rey@upc.edu}
\affiliation{Departament de Física, Universitat Politècnica de Catalunya, \\ Barcelona 08034, Spain.}

\date{\today}

\begin{abstract}
Depending on their mechanism of self-propulsion, active particles can exhibit a time-dependent, often periodic, propulsion velocity. 
The precise propulsion velocity profile determines their mean square displacement and their effective diffusion coefficient at long times. 
Here we demonstrate that any periodic propulsion profile results in a larger diffusion coefficient than the corresponding case with constant propulsion velocity. We investigate in detail the case of periodic exponentially decaying velocity pulses, expected in propulsion mechanisms based on sudden absorption of finite amounts of energy. We show both analytically and with numerical simulations that in these cases the effective diffusion coefficient can be arbitrarily enhanced with respect to the case with constant velocity equal to the average speed. Our results may help interpret in a new light observations on the diffusion enhancement of active particles. 
\end{abstract}
\maketitle
\section{Introduction}
Systems of self-propelled micro/nanoparticles in fluids are currently the focus of considerable attention.
These comprise swimmers which cyclically deform in a non-reciprocal manner (be it by internal consumption of energy or by external actuation) and particles which propel due to self-phoretic mechanisms~\cite{Bechinger2016}. 
Understanding the dynamics of self-propelling particles at the micro and nanoscales is essential to explain a wide range of biological processes, including cellular communications and transport~\cite{Elgeti2015}, or the behavior of catalytic enzymes~\cite{Ghosh2021} among others. The interest in these systems is also due to their disruptive potential applications in different fields such as medicine or biotechnology, since they could be used as efficient drug-delivery vectors~\cite{Ebbens2010,Patra2013, Xu2017}, as non-invasive microsurgery vehicles~\cite{Xu2017}, or in water purification devices~ \cite{Li2014}. 
Furthermore, such systems present fundamental challenges, both in connection to their propulsion mechanisms and their non-equilibrium collective behavior~\cite{Bechinger2016,Cates2015}. 

In particular, numerous works have been devoted to investigate the statistical dynamics of self-propelled particles~\cite{Bechinger2016,Cates2015, Lauga2011, Babel2014}. Being a system intrinsically out of equilibrium, understanding the dynamical behavior of active particles constitutes a great challenge, even in the dilute regime. In this endeavor, the vast majority of studies consider models with constant propulsion velocities. For example, one of the most studied descriptions of self-propelling micro/nanoparticles is given by the active brownian particle (ABP) model. This model describes the dynamics of self-propelled particles in the overdamped regime, often assuming a constant propulsion velocity, and under the action of  stochastic forces of thermal origin which affect both the translational and orientational degrees of freedom of the swimmer~\cite{Romanczuk2012}. Despite its simplicity, the ABP model contains essential ingredients of self-propulsion and has led to a wealth of novel interesting behaviors of active particles ~\cite{Cates2015, Fily2012, Redner2013, Digregorio2018}. 
In particular, the ABP model reproduces the different dynamical regimes in the mean square displacement (MSD) observed for non-interacting self-propelling particles, which becomes diffusive in the long time limit with an effective diffusion coefficient $D_\mathrm{eff}$ which can be orders of magnitude larger than the normal passive diffusivity~\cite{Romanczuk2012}.

Nevertheless, there exist a number of reasons to investigate the effect of a time-dependent propulsion velocity on the dynamics of active particles: (i) Measurements on swimming microorganisms reveal that their propulsion velocity follows an approximately periodic --not constant-- profile~\cite{Friedrich2010, Polin2009}. (ii) In systems of self-phoretic particles, temporal inhomogeneities in the environment (in solute concentration for self-diffusiophoresis; of temperature for self-thermophoresis) can lead to time-dependent propulsion velocities. (iii) In addition, in self-phoretic systems which are externally activated (through, e.g. electromagnetic radiation) time-dependent velocity profiles can be prescribed at convenience~\cite{Jiang2010, Buttinoni2012, Dietrich2020}. (iv) Time-dependent propulsion velocity profiles could also appear in particles propelled by chemical reactions catalyzed at asymmetric surfaces as a result of reactive momentum transfer\cite{Eloul2020}. (v) A propulsion velocity which decays in time can also be expected in vibrationally excited molecules as they dissipate the excess energy anisotropically into the surrounding solvent~\cite{Jurado2023}.

Indeed, the propulsion through time-dependent velocities can lead to distinct and interesting dynamical behaviors. Some of such effects were studied already in earlier works. 
In Ref.~\cite{Lauga2011}, Lauga demonstrated that micro/nanoswimmers undergoing periodic reciprocal motion (with back and forth propulsion velocity) may present enhanced diffusivities above the normal passive diffusion due to the effect of rotational diffusion. This result thus implies that Purcell's scallop theorem \cite{Purcell1977} cannot be extended to lengthscales where the effect of the fluctuating environment is of relevance. Babel and co-workers investigated the statistical dynamics of two-dimensional swimmers with three different time-dependent self-propulsion velocity profiles, unveiling a complex and qualitatively different behavior from the constant velocity case~\cite{Babel2014}. Other works have analyzed the effects of self-propulsion velocities undergoing stochastic transitions between different dynamical states ~\cite{Krasky2018, Jamali2020}, or the effect of non-thermal noise in the propelling velocity of the swimmer~\cite{Thiffeault2022}.

In this article we consider swimmers self-propelled by general time-dependent velocities in a fluctuating environment through an extension of the active brownian particle (ABP) model. We provide expressions to calculate the experimentally accessible MSD and the effective diffusion coefficient ($D_\mathrm{eff}$) for a general velocity profile. We then present a general expression for the time-dependence of the MSD and $D_\mathrm{eff}$ for any periodic propulsion velocity profile in terms of the parameters of its Fourier series expansion.  
From this expression we show that any periodic propulsion profile always results in a larger diffusion coefficient than the corresponding case with constant propulsion velocity (equal to the average velocity).
We also apply such formalism to derive explicit solutions for the MSD and $D_\mathrm{eff}$ for several time dependent velocity profiles, including results for a harmonic time-dependent profiles and periodic profiles of exponentially decaying pulses. In particular, we demonstrate both analytically and with numerical simulations that in the case of periodic exponentially decaying velocity pulses $D_\mathrm{eff}$ can be substantially enhanced with respect to the case with constant velocity for cases where both the time of rotational diffusion and the decay time of the velocity pulse are smaller than the period of the pulses. 

The rest of the article is organized as follows: in  section \ref{sec:1D} we introduce the basic model and formalism to investigate the case of the one dimensional swimmer under an arbitrary time-dependent propulsion velocity profile; as an example of application, results for harmonic profiles are obtained. In section \ref{sec:higherdims}, the formalism is extended to higher dimensions, the harmonic velocity profile revisited, and a general solution for periodic profiles in terms of Fourier series expansion is discussed. In section \ref{sec:exp} the particular case of periodic profiles of exponentially decaying pulses is thoroughly investigated using analytical closed-forms and numerical simulations. This analysis allows us to identify regions in the parameter space where the effective diffusion is significantly enhanced with respect to the corresponding constant velocity profile case.





\section{One dimensional swimmer}\label{sec:1D}
The one dimensional case, thanks to its simplicity, serves to highlight the fundamental issues (both physical and mathematical) involved in the study of a swimmer subject to time dependent propulsion. As is well known, in the low Reynolds number regime friction dominates the dynamics, and the swimmer can be described as an overdamped brownian particle subject to an additional velocity drift~\cite{Hagen2011,Coffey2012}, representing propulsion, and which here will be considered to be time dependent
\begin{equation}
    \dv{x}{t} = \sqrt{2D_T}\zeta_x + v\qty(t) ,
\end{equation}
with the random noise ($\zeta_x$) being characterized by a zero average and delta correlation function (so that $\int_0^t \dd{\uptau}\langle\zeta_x\qty(t)\zeta_x\qty(\uptau)\rangle = 1/2$).

This equation can be formally integrated as
\begin{equation}
    \Delta x\qty(t) = \int_0^t \dd{\uptau}\qty[\sqrt{2D_T}\zeta_x + v\qty(\uptau)]\rm,
\end{equation}
from which the time derivative of the square of the displacement from equilibrium follows
\begin{equation}
    \frac{1}{2}\dv{ \qty[ \Delta x^2 ] }{t}= \Delta x\dv{x}{t}   
    = 2D_T\int_0^t \dd{\uptau}\zeta_x\qty(t)\zeta_x\qty(\uptau)
    +     \sqrt{2D_T}\zeta_x\qty(t)\int_0^t\dd{\uptau}v\qty(\uptau)
    + v\qty(t)\sqrt{2D_T}\int_0^t \dd{\uptau}\zeta_x\qty(\uptau)
    + v\qty(t)\int_0^t\dd{\uptau}v\qty(\uptau)\rm.
\end{equation}

To obtain the mean square displacement (MSD) we perform, as usual, an average over realizations of the random noise (denoted by the subscript $\zeta$), which results in a substantial simplification
\begin{equation}\label{eq:msd1d}
    \dv{\langle \Delta x\rangle^2_{\zeta}}{t} = 2D_T + 2v\qty(t)\int_0^t \dd{\uptau} v\qty(\uptau)\rm .
\end{equation}

The constant velocity ($v(t) = v $) case can be immediately solved for, 
\begin{equation}
\left\langle \Delta x^2(t) \right\rangle = 2 D_T t + v^2 t^2    ,
\end{equation}
displaying diffusive dynamics at short times followed by ballistic (super-diffusive) motion at long times.

Time varying velocity is our focus of interest and here a careful consideration of averaging needs to be made. In principle one could insert a given velocity profile (defined in the interval $[0,\infty)$) directly into Eq. \ref{eq:msd1d}, but this choice will generally produce formulas dependent on initial conditions, with an unnecessary increase in complexity. Here we perform an average over origins along the swimmer's trajectory, in line with a common experimental measurement procedure (equivalent to an averaging over independent particle trajectories). No such sampling was required in the constant velocity case (or for a simple brownian particle), as all origins along the trajectory are indistinguishable, but now each origin (or particle) may be characterized by a distinct propulsion velocity. This additional average (denoted with $<>_v$) will only affect the second term in Eq. \ref{eq:msd1d}
\begin{equation}\label{eq:1d-vmean}
    \dv{\langle \Delta x\rangle^2_{\zeta v}}{t} = 2D_T + 2\left\langle v\qty(t)\int_0^t \dd{\uptau} v\qty(\uptau)\right\rangle_v\rm .
\end{equation}

\begin{figure}[H]
    \centering
    \includegraphics[width=0.48\textwidth]{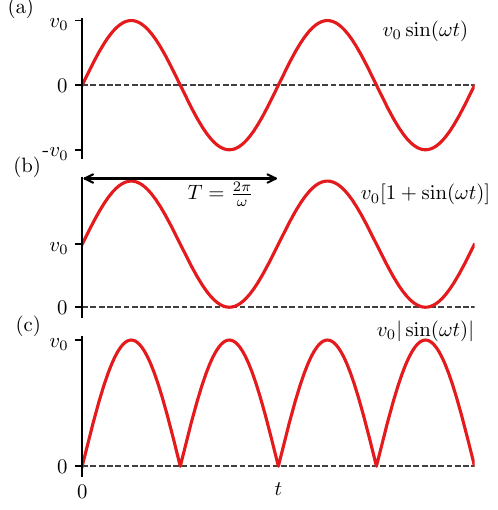}
    \caption{Representation of the different sinusoidal velocity profiles studied in this work. (a) zero mean velocity harmonic profile; (b) shifted harmonic profile with non-zero mean velocity;(c) absolute value of sine function (abs-sin in text). \label{fig:sine_profiles}
    }
\end{figure}

A more compact form can be derived for the specific case of periodic velocity profiles, by noting that the average over origins can be restricted to shifts along a single period ($T$), that is 
\begin{equation}\label{eq:dmsd-1d-a}
    \dv{\langle \Delta x\rangle^2_{\zeta v}}{t} = 2 D_T + 2\frac{1}{T}\int_0^T \dd{s} \qty{  v(t-s) \int_0^t v(\uptau -s)\dd{\uptau} }\rm, 
\end{equation}
or alternatively 
\begin{equation}\label{eq:dmsd-1d-b}
    \dv{\langle \Delta x\rangle^2_{\zeta v}}{t} = 2 D_T + 2\int_0^t \dd{\uptau}\qty{ \frac{1}{T}\int_0^T \dd{s}   v(t-s) v(\uptau -s) }\rm.
\end{equation}

We start with their application to two periodic velocity profiles of particular interest (and to which we will return for comparison when addressing the multidimensional case). First, we consider a harmonic profile with zero mean ($v(t)= v_0 \sin(\omega t + \theta_0)$), displayed in Fig. \ref{fig:sine_profiles}(a)), for which integration of Eq. \ref{eq:dmsd-1d-a} (or Eq. \ref{eq:dmsd-1d-b}) yields
\begin{equation}\label{eq:msd-1d-harm}
\left\langle \Delta x^2\qty(t) \right\rangle = 2 D_T t + \frac{v_0^2}{\omega^2}\qty[1- \cos(\omega t)]    ,
\end{equation}
a result which in principle was to be expected: a propulsion which periodically switches its direction has a null impact on diffusion at long times (notice that with the additional averaging over origins the results are independent of any initial phase, $\theta_0$; see section I.A in Supplementary Information, SI). The second case corresponds to harmonic propulsion as well, but now described by the profile $v(t) = v_0 \qty[1 + \sin(\omega t)]$, i.e. a sequence of positive definite pulses with average $\langle v\rangle =v_0$ (see Fig. \ref{fig:sine_profiles}(b). We obtain
\begin{equation}
\left\langle \Delta x^2\qty(t) \right\rangle = 2 D_T t + v_0^2 t^2  + \frac{v_0^2}{\omega^2}\qty[1- \cos(\omega t)] ,
\end{equation}
i.e., the simple addition of the result for constant velocity and that for (alternate) sinusoidal propulsion. The crucial point here is the null effect of propulsion velocity fluctuations (in both harmonic cases), it is solely the average velocity which matters at long times. No benefit is obtained from resorting to a time dependent propulsion velocity, compared to constant velocity propulsion (as far as it equals the average for time dependent propulsion). This statement will change for higher dimensions, dramatically if the shape of the pulses is conveniently tailored.

\section{Higher dimensions}\label{sec:higherdims}

\subsection{General formalism}\label{sec:gen-form}
The basic change is the introduction of rotational diffusion of the direction of propulsion in addition to center of mass (CM) overdamped dynamics. The standard dynamical equations in two dimensions are\cite{Hagen2011,Coffey2012}
\begin{eqnarray}
    &&\dv{x}{t} = \sqrt{2D_T}\zeta_x + v\qty(t)\cos\theta\rm,\\
    &&\dv{y}{t} = \sqrt{2D_T}\zeta_y + v\qty(t)\sin\theta\rm,\\
    &&\dv{\theta}{t} = \sqrt{2D_R}\zeta_\theta\rm .
\end{eqnarray}
We note that now there are several noise processes ($\{\zeta_x,\zeta_y,\zeta_\theta\}$), and two diffusion coefficients (corresponding to CM translations, $D_T$, and to rigid body rotations, $D_R$). Again, the time dependence of the propulsion velocity has been made explicit. 

The equations for the CM dynamics can be written in a more compact vectorial notation (valid for any dimension) as
\begin{equation}\label{eq:predebye}
\dv{\va{r}}{t} = \sqrt{2D_T}\va{\zeta} + v\qty(t)\vu{e},
\end{equation}
where $\hat{e}$ stands for the unit vector pointing along the direction of propulsion, and which itself is subject to the typical Debye dynamics of a damped spherical rotator, characterized by a correlation function \cite{Caillol2004}
\begin{equation}\label{eq:debye}
 \left\langle \vu{e}\qty(t) \cdot \vu{e}\qty(\uptau)\right\rangle_\theta = e^{-\frac{t-\uptau}{\tau_r}}.
\end{equation}
The new parameter $\tau_r$ is directly related to the rotational diffusion coefficient, although with an additional dependence on the dimension $d$ ($\tau_r \equiv \frac{1}{(d-1) \cdot D_R}$). It is therefore convenient to use $\tau_r$ instead of $D_R$ when possible in order to obtain formulas as general as possible.

At this point it is straightforward to derive (along the lines followed in the one-dimensional case) the corresponding expression for the MSD and to perform the averages over random noises (subscript $\zeta$ for center of mass motion, $\theta$ for rotations),
\begin{equation}
    \dv{\langle \Delta\va{r}^2\rangle_{\zeta \theta}}{t} = 2 d\cdot D_T + 2 v\qty(t) \int_0^t \dd{\uptau} v\qty(\uptau)\left\langle \vu{e}\qty(t) \cdot \vu{e}\qty(\uptau)\right\rangle_\theta  \rm.
\end{equation}

Inserting the rotational time correlation relaxation function (Eq. \ref{eq:debye}), 
\begin{equation}
    \dv{\langle \Delta\va{r}^2\rangle_{\zeta \theta}}{t} = 2 d\cdot D_T + 2  v\qty(t)e^{-t/\tau_r} \int_0^t \dd{\uptau} v\qty(\uptau)e^{\uptau/\tau_r}  \rm.
\end{equation}
an expression that underlines the trivial difference between dimensions ($d$ prefactor of the translational diffusion coefficient, $D_T$), and which includes the one dimensional case as well (according to the definition of $\tau_r$ the exponentials cancel out, and Eq. \ref{eq:msd1d} is recovered). 

Following the one-dimensional example, we introduce an additional average over origins, which results in
\begin{equation}\label{eq:dbasic}
    \dv{\langle \Delta\va{r}^2\rangle}{t} = 2 d\cdot D_T + 2\left\langle  v\qty(t)e^{-t/\tau_r} \int_0^t \dd{\uptau} v\qty(\uptau)e^{\uptau/\tau_r} \right\rangle_{v} \rm,
\end{equation}
the basic expression that will be used throughout.

For constant speed we trivially recover the standard formula\cite{Bechinger2016}
\begin{equation}\label{eq:msd-v-standard}
    \dv{\langle \Delta \va{r}^2\rangle}{t} = 2 d\cdot D_T + 2 v^2 \tau_r\left(1 - e^{-t/\tau_r}\right) .
\end{equation}
The exponential transient, with a characteristic time equal to $\tau_r$, implies that the dynamics (contrary to the, ballistic, one dimensional behaviour) is diffusive $\qty(\left\langle  \left[ \Delta \vec{r}(t) \right]^2\right\rangle \xrightarrow{t \rightarrow \infty } 2 d D_\mathrm{eff} t)$, 
\begin{equation}\label{eq:d-vconst}
    D_\mathrm{eff}  = D_T +  \frac{v^2 \tau_r}{d}\equiv D_T + D_\mathrm{ex} ,
\end{equation}
where the excess part ($D_\mathrm{ex}$) stems from rotation, i.e. although rotation substantially dampens the MSD (compared with the quadratic superdiffusive regime in 1-d), diffusion is still enhanced with respect to the free case. This simple expression can be rewritten into a form which will be particularly clarifying when it comes to analyze the results for time dependent propulsion. It involves the P\'eclet number ($\mathrm{Pe}$) defined \cite{Bechinger2016} 
\begin{equation}\label{eq:pe}
\mathrm{Pe} \equiv \frac{\left| v \right|}{\sqrt{2 D_T D_R}}    ,
\end{equation}
where $\left| v \right|$ denotes the absolute value of the propulsion velocity (i.e. speed). 

In terms of $\mathrm{Pe}$, Eq. \ref{eq:d-vconst} is simply expressed as
\begin{equation}\label{eq:d-pe}
    D_\mathrm{eff}  = D_T\left[1 + \frac{2}{d(d-1)} \mathrm{Pe}^2 \right] ,
\end{equation}
with an appealing interpretation: $\mathrm{Pe}$ can be viewed as encapsulating, in a single dimensionless parameter, the enhancement to free diffusion which stems from the interplay between propulsion and rotation. We note that for $d=2$ Eq.~\ref{eq:d-pe} yields the simple relation $D_\mathrm{eff}  =D_T\left[1 + \mathrm{Pe}^2 \right]$, and $D_\mathrm{eff}  =D_T\left[1 + \mathrm{Pe}^2/3 \right]$ for $d=3$. For the sake of generality the explicit dimension dependence will be kept within (although the simulations to be reported will correspond in all cases to $d=2$, and the results compared with the corresponding formulas).

Finally, as it was the case in 1-d, a compact formulation results for periodic velocity profiles
\begin{eqnarray}
   \left\langle  v\qty(t)e^{-t/\tau_r} \int_0^t \dd{\uptau} v\qty(\uptau)e^{\uptau/\tau_r} \right\rangle_{v} &&= \frac{1}{T} \qty{ \int_0^T \dd{s} v(t-s)e^{-t/\tau_r} \int_0^{t} v(\uptau -s)e^{\uptau/\tau_r}\dd{\uptau}}= \label{eq:v1}\\
  &&= \frac{1}{T}  e^{-t/\tau_r} \int_0^t\dd{\uptau} \qty{\int_0^T \dd{s} v(t-s)v(\uptau -s)}e^{\uptau/\tau_r} \label{eq:v2},
\end{eqnarray}
where the only formal change, with respect to one dimension, is the introduction of exponential corrections resulting from orientational relaxation.

\subsection{Sinusoidal propulsion with zero average velocity}
We now revisit the first time dependent case analyzed in 1-d, namely $v(t)= v_0 \sin(\omega t)$. This velocity profile can be readily integrated after inserting it in Eq.~\ref{eq:v1} (or Eq.~\ref{eq:v2}). The full MSD reads,
\begin{widetext}
\begin{equation} \label{eq:d-sin2d}
\left\langle \left[ \Delta \vec{r}(t) \right]^2\right\rangle = 2d \left[ D_T + \frac{v_0^2 \tau_r}{2d \qty[1 + (\tau_r \omega)^2]}\right] t    + \frac{v_0^2(\omega^2 - \tau_r^{-2})}{(\omega^2 + \tau_r^{-2})^2} -
\frac{v_0^2 e^{-t/\tau_r}}{(\omega^2 + \tau_r^{-2})^2}\left[(\omega^2 - \tau_r^{-2})\cos(\omega t)+2 \frac{\omega}{\tau_r}\sin(\omega t)\right].
\end{equation}    
\end{widetext}

\begin{figure}[H]
    \centering
    \includegraphics[width=0.48\textwidth]{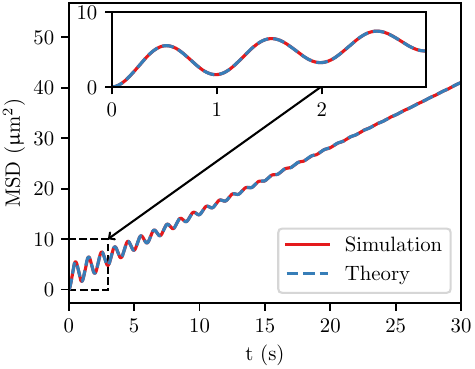}
    \caption{Mean Square Displacement for a (zero average) sinusoidal velocity profile obtained via simulation (red solid) and via Eq.~\ref{eq:d-sin2d} (blue dashed). The inset shows the short time behaviour of the MSD. The simulation data lines have an error smaller than the thickness of the line. Simulations were performed, in 2-d, with $D_T = 0.22\:\upmu\mathrm{m}^2/s$, $D_R = 0.16\:\mathrm{rad}^2/s$, $v=10\:\upmu\mathrm{m}/s$ and $\omega=2\pi\:\mathrm{rad}/s$, corresponding to $\avgPe\simeq 24$ and $F\simeq 0$ (see text). \label{fig:msd_sin}}
\end{figure}

The last two terms correspond to the transient, characterised by a time scale $\tau_r$, as was also the case for constant propulsion velocity. Figure~\ref{fig:msd_sin} illustrates the transition from transient (inset) to stationary behaviour for a particular instance. The analytical expression perfectly matches numerical simulation (Supplementary Information, SI, contains a summary of the numerical techniques employed).

The first term in Eq.~\ref{eq:d-sin2d} defines an effective diffusion coefficient ($D_\mathrm{eff}= D_T + \frac{v_0^2 \tau_r}{2d (1 + (\tau_r \omega)^2)}$) valid in the long-time limit,  which was already obtained in Ref.~\onlinecite{Lauga2011}. While we are dealing with a simple case (sinusoidal), it has broad implications.~\cite{Lauga2011} Namely, the scallop theorem~\cite{Purcell1977} needs to be qualified if orientational fluctuations are not negligible: a time reversible propulsion sequence can actually enhance the underlying diffusional dynamics (described by $D_T$). Finally, the enhancement of diffusion found in two/three dimensions contrasts with the corresponding null effect in one dimension for a zero mean propulsion profile such as this one.

\subsection{Average P\'eclet number}
The effective diffusion coefficient can also be expressed in terms of the P\'eclet number, if its definition is extended to time dependent velocity profiles. We define the \textit{average} P\'eclet number as the simple average of Eq.~\ref{eq:d-pe}
\begin{equation}
\avgPe \equiv \frac{\langle \qty|v| \rangle}{\sqrt{2 D_T D_R}}    .
\end{equation}

For the zero average harmonic velocity case, $\langle \qty|v| \rangle = 2 v_0/\pi$ and therefore $\avgPe^2 = 2 v_0^2/\pi^2 D_T D_R $. The effective diffusion coefficient can thus be written 
\begin{equation}\label{eq:pe-f}
    D_\mathrm{eff}   = D_T\qty[1 + \frac{2}{d(d-1)}\avgPe^2F(\tau_r')]\rm,
\end{equation}    
where a new function has been introduced, 
\begin{equation}\label{eq:f-sin}
F(\tau_r') \equiv \frac{\pi^2}{8}\frac{1}{1+(\tau_r \omega)^2}\rm.
\end{equation}

Eq.~\ref{eq:pe-f} is highly similar to Eq.~\ref{eq:d-pe}, with the only addition of a multiplicative functional correction ($F$) to the average P\'eclet number. While $\avgPe$ encapsulates the average enhancement of diffusion, the new multiplicative factor ($F$) reflects the correction to this first, constant velocity (equal to the average speed of the profile), approximation. This will be the framework adopted within in order to help compare more realistic velocity profiles. While the corresponding diffusion coefficients will display a substantially higher mathematical complexity, when expressed in terms of Eq.~\ref{eq:pe-f}, the only difference with respect to the present case will be that the ``fluctuation correction factor'' ($F$) is multidimensional (with the dimensionless variables being the set of, scaled, natural timescales of the propulsion velocity profile). The dimensionless nature of all the functions and variables involved, will make the analysis valid for any system/scale (as long as the underlying dynamical equations apply).  

In the following, propulsion profiles with the same $\avgPe$ will be compared, and thus the focus will be (according to Eq.~\ref{eq:pe-f}) on comparison of the respective correction factors ($F$). This procedure though needs to be exercised with care. In most cases the comparison will be between periodic profiles which are indistinguishable in terms of their absolute value (see for instance panels (a) and (c) in Fig.~\ref{fig:sine_profiles}). Their average P\'eclet numbers are mathematically identical, and therefore it is fair to directly compare their respective correction factors ($F$). A different kind of comparison will be between profiles with different functional dependence, a basic one being between constant propulsion and a time varying profile, both with the same $\avgPe$, which in principle constitutes a natural approach. However, since the goal is to discern which one is optimal in terms of diffusion enhancement, one would actually want to compare profiles characterized by the same energy expense, a criterion in principle not necessarily identical to the equality of average P\'eclet numbers. Both criteria are equivalent though under the reasonable assumption of a linear dependence between average speed (i.e. $\avgPe$) and average power deposited on the swimmer. Indeed, such linear relationship has been shown to hold to a very good approximation (Ref. \onlinecite{Calero2020}, Fig. 2) for nanometric swimmers in water (functionalized fullerene in bulk liquid water or inside a carbon nanotube), although the range of validity will generally need to be checked for each particular system. 

\subsection{Fourier solution}
Let us consider now the effects on the dynamics of ABPs propelled by the general subclass of velocity profiles which exhibit periodicity in time. We can express such velocity profiles with the corresponding Fourier series expansion

\begin{equation}\label{eq:four1}
    v\qty(t) = \sum_{n=-\infty}^{\infty}c_n e^{in\omega t} \qq{with} \omega\equiv\frac{2\pi}{T} ,
\end{equation}
where
\begin{equation}\label{eq:four2}
    c_n \equiv\frac{1}{T}\int_0^T \dd{t} v\qty(t)e^{-in\omega t}\rm.
\end{equation}

When substituted into Eq.~\ref{eq:dbasic} we obtain, after some manipulation (see SI), an expression for the MSD that applies to \textit{any} periodic propulsion velocity profile
\begin{widetext}
\begin{equation}\label{eq:dfourier-general}
\dv{\langle \Delta \va{r}^2\rangle}{t} = 2d\qty[D_T + \frac{c_0^2 \tau_r}{d} + \frac{2}{d}\sum_{n=1}^\infty \qty|c_n|^2\frac{\tau_r}{1+\qty(n\omega \tau_r)^2}]
-2c_0^2 \tau_r e^{-t/\tau_r} 
+4\sum_{n=1}^\infty \qty|c_n|^2\frac{e^{-t/\tau_r}}{\tau_r^{-2}+\qty(n\omega)^2}\qty[n\omega\sin(n\omega t)-\tau_r^{-1}\cos(n\omega t)]\rm .
\end{equation}
\end{widetext}

From this full analytic expression for the MSD it is immediate to recover its known long time limit~\cite{Lauga2011}, which defines an effective diffusion coefficient  
\begin{equation}\label{eq:d-fourier}
D_\mathrm{eff} = D_T + \frac{c_0^2 \tau_r}{d} + \frac{2}{d}\sum_{n=1}^\infty \qty|c_n|^2\frac{\tau_r}{1+\qty(n\omega \tau_r)^2}\rm.
\end{equation}
Although it certainly was to be expected, this expression clearly shows~\cite{Lauga2011} how any propulsion velocity will result in an enhancement of diffusion with respect to free center of mass diffusion (i.e. $D_\mathrm{eff} > D_T$), as implied by the positive definite nature of the corrections (second and third terms). Note that despite Eq.~\ref{eq:dfourier-general} is obtained by averaging over the particles' initial phase, the resulting expression for the effective diffusion coefficient, Eq.~\ref{eq:d-fourier}, is valid for any periodic propulsion velocity profile with no necessity of averaging over the time origins of the particles (see section I.C in SI).

It may have gone unnoticed, and this is central to the present work, that Eq. \ref{eq:d-fourier} also implies that any periodic propulsion profile will always be more efficient than constant propulsion velocity (equal to the average velocity). This results from noting that, for any periodic profile, $c_0 = \langle v\rangle$, so the first two terms ($D_T + \frac{c_0^2 \tau_r}{{d}}$) are identical to the diffusion coefficient for the corresponding (constant) average velocity (Eq.~\ref{eq:d-vconst}). Since the third term (infinite sum) in Eq.~\ref{eq:d-fourier} is positively defined, regardless of the shape of the (periodic) velocity profile, a diffusion enhancement will necessarily result with respect to constant velocity. 

This observation for the long time dynamics contrasts with the transient dynamics previously described, its associated time scale ($\tau_r$) does not differ from what was previously found for constant velocity (Eq.~\ref{eq:msd-v-standard}), or for zero average harmonic propulsion velocity (Eq.~\ref{eq:d-sin2d}). It is evident now that these two examples were just instances of a general result: irrespective of the functional form and/or period of the (periodic) velocity propulsion imparted on the swimmer, the duration of the transient is uniquely determined by rotational diffusion. 

To summarize, while resorting to periodic propulsion will not result in any remarkable effect on the transient lifetime, for the effective diffusion coefficient an advantage is guaranteed in all cases with respect to a constant velocity equal to the mean velocity of the profile. 

\begin{figure}[H]
    \centering
    \includegraphics[width=0.48\textwidth]{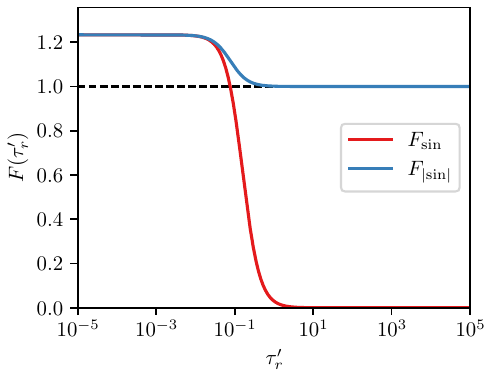}
    \caption{Fluctuation correction factors for a null mean velocity sine profile (red) and absolute value sine profile (blue). The corresponding constant velocity value is shown as a dashed black line.\label{fig:sin_absin_1}}
\end{figure}

We start exploring the scope of this treatment by completing the analysis for harmonic profiles. As a first application we discuss the profile given by the absolute value of the sinus function (\emph{abs-sin} from here on, displayed in Fig.~\ref{fig:sine_profiles}(c)) since it cannot be solved by direct integration of Eq.~\ref{eq:v1} (or Eq.~\ref{eq:v2}). Using Eq.~\ref{eq:d-fourier}, the following effective diffusion coefficient results (see SI) for the abs-sin profile

\begin{equation}
D_\mathrm{eff}  =  D_T + \frac{4 v_0^2 \tau_r}{{d}\pi^2} \left[ {1}+{2}\sum_{n= 1}^\infty\frac{1}{(1- 4 n^2)^2\qty[1+(2n\omega \tau_r)^2]} \right] ,
\end{equation}
which when rewritten in the generic form $D_\mathrm{eff} =  D_T \left[ 1+ {\frac{2}{d(d-1)}}\left\langle \mathrm{Pe}\right\rangle^2 F(\tau_r')\right]$, defines the corresponding fluctuation correction factor for the abs-sin profile
\begin{equation}\label{eq:f-abssin}
 F(\tau_r')=  1+ 2\sum_{n = 1}^\infty \frac{1}{(1- 4 n^2)^2 \qty[1+(4 \pi n \tau_r')^2]} ,
\end{equation}
displayed in Fig.~\ref{fig:sin_absin_1}, alongside the corresponding curve for a zero average sinusoidal propulsion (Eq.~\ref{eq:f-sin}). Here, we have defined the dimensionless parameter $\tau_r' \equiv \omega \tau_r$.

As it was to be expected from the previous discussion, the abs-sin fluctuation correction factor (blue curve in Fig.~\ref{fig:sin_absin_1}) is always larger or equal than what would result from constant velocity (equal to the average speed) propulsion (horizontal dashed line). This enhancement, though, is rather modest, with a maximum increase of $\simeq 20$\% (from Eq.~\ref{eq:f-abssin} it is immediate to see that $F(0) = \pi^2/8 \simeq 1.2 $). This small advantage appears when rotation grows faster, with the transition being located at $\tau_r'\simeq 1$ (i.e. $\tau_r \simeq T$).

Fig.~\ref{fig:sin_absin_1} also displays the result for purely sinusoidal (zero mean) propulsion (red curve), corresponding to our previous result in Eq.~\ref{eq:f-sin}. For slow rotational diffusion, and/or large period (i.e. $\tau_r' \rightarrow \infty$), we have $F(\tau_r')\rightarrow 0$. This is consistent with the notion that if the swimmer does not rotate, the dynamics is in fact indistinguishable from 1-d, where we found no effect from a harmonic propulsion velocity (see Eq.~\ref{eq:msd-1d-harm}). A dramatic change though occurs for $\tau_r'\simeq 1$: as rotation speeds up (and/or the period gets larger), this propulsion profile can become as efficient as the corresponding constant velocity case (dashed line) and even outperform it. Ultimately it becomes as efficient as the abs-sin profile (overlap of blue and red curves, the limit for $\tau_r'\rightarrow 0 $ is exactly the same as found for abs-sin, $\pi^2/8$, as can be trivially checked from Eq.~\ref{eq:f-sin}). The explanation is rather simple, for fast rotational diffusion one-way or alternate propulsion directions become indistinguishable. 

We could summarize this Section by noting that, while harmonic fluctuations have been the center of attention when it comes to velocity variations, their contribution (to long time diffusion) is at best marginal with respect to constant velocity, in any dimension. This adds to
an equal time-dependence of the MSD transient towards its long time behavior, which applies to \textit{any} periodic profile (harmonic or not).  

These considerations are suggestive that there might be no point in trying to exploit a time dependent propulsion velocity. As noted in the Introduction though, for a wide set of systems a harmonic modelization is probably not realistic enough, if propulsion is to proceed through short bursts. These are associated in most cases to swimmer-solvent fast energy transfers which can be produced, as described in the Introduction as well, by a variety of mechanisms that will depend on the swimmer's length scale~\cite{Calero2020, Eloul2020, Jurado2023}. The consideration of short-lived pulses (when compared with the period with which they are applied), introduces an additional time scale: while for the harmonic case the only variable was the ratio between rotational relaxation time and period ($\tau_r' = \tau_r/T$), now the ratio between pulse duration and period will constitute a new (dimensionless) variable to consider. Its impact, which does not seem to have been addressed before, is the focus of next Section.

\section{Exponential profile}\label{sec:exp}
A propulsion velocity profile made of a train of decaying exponentials is a natural choice, due to both physically motivated reasons (an exponential decay is to be expected for absorption of finite energy bursts), and to mathematical simplicity (it only requires the introduction a single new variable, its lifetime $\tau$, and allows for explicit solutions, as will be shown). 

\begin{figure}[H]
    \centering
    \includegraphics[width=0.48\textwidth]{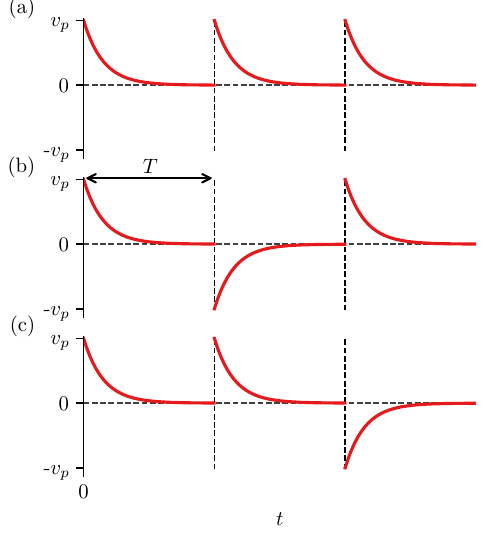}
    \caption{Exponential propulsion velocity profiles for: (a) same sign velocity kicks ($++$); (b) alternate sign ($+-$); (c) and random sign ($\mathrm{rnd}$). \label{fig:exp_profiles}}
\end{figure}

We will thus analyze velocity profiles constituted by an equally time shifted (by a time $T$) sequence of signed instances of the exponential
\begin{equation}
    v\qty(t) = v_p e^{-t/\tau}\rm,
\end{equation}
where $v_p$ stands for the absolute value of the peak velocity (the subscript is added in order to avoid confusion in the results to be reported). Three different sequences will be considered (sketched in Fig. \ref{fig:exp_profiles}): 
\begin{enumerate}[(a)]    
    \item Same sign, denoted by the ($++$) symbol,
    \item Alternating sign ($+-$),
    \item Random sign ($\mathrm{rnd}$).
\end{enumerate}

As has been the case for harmonic profiles, where sequences with zero and non-zero mean were studied, the last two cases are representative of sequences with null average velocity ((b) and (c)), while the first one corresponds to an instance for which propulsion direction is maintained between pulses (sequence (a)). As the functional dependence for each exponential (``kick'') is identical, all instances automatically have the same average speed and P\'eclet number, so that only the fluctuation correction factors need be compared. Finally, we note that the third sequence is characterized by the lack of periodicity, the first and only example with this characteristic, although it displays time regularity (equally spaced kicks). 

\subsection{Analytic solution}
We start with the same sign (++) sequence, for which an analytic MSD can be obtained by integrating the general Fourier solution for its derivative (Eq. \ref{eq:dfourier-general}). The following expression results (see SI),
\begin{widetext}
\begin{equation}
\langle \qty[\Delta \va{r}(t)]^2\rangle = {2d}\qty[D_T + \frac{\left\langle v\right\rangle^2 \tau_r}{{d}} + {\frac{2}{d}}\left\langle v\right\rangle^2 \tau_r\sum_{n=1}^\infty \frac{1}{\qty[1+\qty(n\omega \tau)^2 ]\qty[1+\qty(n\omega \tau_r)^2]}]t + 
\langle \qty[\Delta \va{r}(t)]^2\rangle_\mathrm{trans}\rm,
\end{equation}    
with the transient part given by
\begin{equation}
\langle \qty[\Delta \va{r}(t)]^2\rangle_\mathrm{trans} =   
2\left\langle v\right\rangle^2 \tau_r^2 \qty[e^{-t/\tau_r}- 1] 
+4 \left\langle v\right\rangle^2 \tau_r^2 \sum_{n=1}^\infty \frac{\qty[\qty(n\omega \tau_r)^2-1]\qty[1-e^{-t/\tau_r}\cos(n\omega t)]-
2 n \omega \tau_r \sin(n\omega t)e^{-t/\tau_r}}{\qty[1+\qty(n\omega\tau)^2 ]\qty[1+\qty(n\omega \tau_r)^2]^2}
\end{equation}
\end{widetext}
and where  the average velocity ($\langle v\rangle$ which, in this case, is also equal to the average speed) is 
\begin{equation}
    \langle v\rangle = \frac{1}{T}\int_0^T \dd{\uptau} v\qty(\uptau)=v_p\frac{\tau}{T}\qty(1-e^{-T/\tau})\rm.
\end{equation}
While it has already been stated with full generality, we note in passing that the transient part decays with a time scale equal to the rotational lifetime, as expected. Regarding our main interest, namely the effective diffusion coefficient, when expressed in terms of the average P\'eclet number (see Eq.~\ref{eq:pe-f}) the following fluctuation correction factor results in
\begin{equation}\label{eq:fpp}
    F^{++}(\tau',\tau_r') = 1+2\sum_{n=1}^\infty \frac{1}{\qty[1+\qty( 2\pi n\tau')^2 ]\qty[1+\qty(2\pi n\tau_r')^2]} \rm .
\end{equation}
As was anticipated, it is now a function of two dimensionless variables ($\tau', \tau_r'$). The first one is related to the scaled propulsion relaxation time ($\tau' \equiv \tau/T$), while the second one corresponds, as for the cases studied so far, to the rotational relaxation time ($ \tau_r' \equiv \tau_r/T$). 

Given the positive definite character of the sum in Eq.~\ref{eq:fpp}, the $F^{++}$ correction (to diffusion enhancement produced by a constant velocity, and reflected in $\avgPe$, see Eq.~\ref{eq:d-pe}) will in all cases be $\geq 1$. This is consistent with the general finding that periodic propulsion is always more efficient (in terms of enhancing diffusion) than constant propulsion (with the same average speed). This explicit expression allows though for a more in-depth analysis by studying some of its limits. If any of the two variables is increased ($\tau'\rightarrow \infty$ and/or $\tau_r' \rightarrow \infty$) the sum will tend to zero and therefore $F^{++}\rightarrow 1$, that is, if the period $T$ is much smaller than any of the relaxation time scales, there is no advantage in time variation. For the $\tau >> T$ case this can be understood considering that the exponential decay will tend to be negligible in between kicks, so that the profile will not differ from constant propulsion. The $\tau_r >> T$ limit is identical with that already found for harmonic pulses, i.e. there is no advantage in velocity variation if rotational diffusion is slow.

It is the opposite limit though which is of particular interest. If the period is much larger than \textit{both} relaxation times (or equivalently $\tau', \tau_r' \rightarrow 0$), both denominators within the sum in Eq.~\ref{eq:fpp} will tend to a unit value, and consequently the sum will \textit{diverge}. The fluctuation correction factor can thus grow indefinitely, in strong contrast with the harmonic case (where a maximum value of $\simeq 1.2$ was found). We consider this a key result of this work, as it shows that the diffusion enhancement caused by a constant velocity can be largely exceeded if a sequence of exponential kicks is used instead. It is important to note that this was a double limit and it is not possible to disentangle at the present stage the separate role of each variable ($\tau', \tau_r'$), so we postpone a more in depth analysis to Section~\ref{sec:param_maps}, including consideration of the meaning of the apparent divergence. 

A similar calculation for an alternate sign (+-) sequence yields (see SI)
\begin{eqnarray}\label{eq:fpm}
    F^{+-}(\tau',\tau_r') &=& 2 \qty(\frac{1+ e^{-1/\tau'}}{1-e^{-1/\tau'}})^2 \sum_{n=1,3,...}^\infty \frac{1}{\qty[1+\qty(n \pi \tau')^2 ]\qty[1+\qty(n \pi \tau_r')^2]} \rm.
\end{eqnarray}
The unit value before the sum found for Eq.~\ref{eq:fpp} is now missing, so the correction tends to zero rather than one for $\tau', \tau_r' \rightarrow \infty$, similarly to the (zero average velocity) sinusoidal case. For $\tau', \tau_r' \rightarrow 0$ though, we still find that $F^{+-}$ diverges. That is, somewhat counter-intuitively, even with alternate sign propulsion the enhancement of diffusion can be made indefinitely larger than that corresponding to constant propulsion (with velocity equal to the average speed). Again, it does not seem possible to disentangle the role of each of the variables with the current analytical expression for $F^{+-}$.

The remarkable characteristics found for the exponential propulsion profiles call for a detailed analysis of the two-dimensional parameter space spanned by the fluctuation correction factor, $F^{++/+-}(\tau',\tau_r')$, including all the corresponding (idealized) limits. It has been already noted though that this is not generally possible with the analytic formulas discussed so far. In addition, while $++$ and $+-$ sequences are intuitively appealing, they represent two extremes amongst an infinite number of possible sequences. A random sequence might be of interest as well, as representative of what lies in between. Unfortunately, the Fourier formalism used so far is unable to treat this case, as it shows no periodicity. Remarkably, the exponential propulsion pulses allow for the derivation of closed-formulas in all three cases, a fact directly related to the kernel in the integral (Eq.~\ref{eq:dbasic}) being also exponential. These alternative formulas do not render the Fourier approach useless though. Having two different, and independent, explicit solutions (analytic and closed-form) allows an exhaustive check of their validity without the need to resort to simulation. Moreover, the Fourier methodology is the only approach which can provide an analytic expression of the full MSD (including transient) for any periodic pulse profile. Finally, while it has been straightforward to infer from the Fourier approach the notion that time variation is always more efficient, this fundamental characteristic would not be obvious from inspection of closed-form formulas for particular profiles.   

\subsection{Closed-form solution}
A detailed derivation of closed-form solutions for the effective diffusion coefficient can be found in the Appendix, for each of the sequences considered. As has already been emphasized, the functional forms being identical, the analysis can be circumscribed  to the fluctuation correction factor, for which all three cases can be merged into the single expression 
\begin{widetext}
\begin{equation}\label{eq:fclosed}
    F(\tau',\tau_r') = \frac{1}{\qty(\tau'-\tau_r')\qty(1-e^{-1/\tau'})^2}\qty[\frac{e^{-2/\tau'}-e^{1/\tau_r'-1/\tau'}}{\tau' + \tau_r'}\tau_r' g(\tau',\tau_r')+ \frac{1-e^{-2/\tau'}}{2}] ,
\end{equation}
\end{widetext}
where a new function has been defined ($ g(\tau',\tau_r')$), which encapsulates the only differences amongst them
\begin{equation} \label{eq:gfactor}
 g(\tau',\tau_r')  =
    \begin{cases}
      \frac{1-e^{1/\tau'}}{1-e^{1/\tau_r'}} & \text{for ++ pulses,}\\
      \frac{1+e^{1/\tau'}}{1+e^{1/\tau_r'}} & \text{for $+-$ pulses,}\\
      e^{\qty(1/\tau'- 1/\tau_r')} & \text{for random pulses}.
    \end{cases}       
\end{equation}

SI contains a numerical comparison of the analytic expressions (Eq.~\ref{eq:fpp} for the $++$ sequence, Eq.~\ref{eq:fpm} for the $+-$ sequence) with the corresponding instance of the closed-form formula (Eq.~\ref{eq:fclosed}), and over several orders of magnitude for each one of the two scaled variables. A perfect match (within machine precision) is found between both explicit solutions and for both types of sequence. For random sequences, lacking an analytic expression, it has been necessary to resort to numerical simulations for representative sample cases, as detailed in SI. The agreement is in all cases compatible with the statistical indeterminacy associated with the limited number of independent trajectories generated. 

\subsection{Parameter space characterization}\label{sec:param_maps}
With the help of the closed-form formulas, it is now straightforward to analyze in full detail the behaviour of $F(\tau',\tau_r')$ within the two dimensional parameter space spanned by its dimensionless variables. In order to gain some insight we start by providing a graphical representation for a necessarily limited portion. Figure~\ref{fig:f_3panel} displays a contour plot for each one of the sequences.  

Despite the mathematical complexity of Eq.~\ref{eq:fclosed} its behavior can be summarized as a monotonic increase of the correction factor with decreasing value of the two variables (for all three sequences). A first, qualitative, comparison of Fig.~\ref{fig:f_3panel}(a) and Fig.~\ref{fig:f_3panel}(b) shows that $++$ sequences are, as expected somewhat more efficient than $+-$ sequences and, remarkably, comparison of Fig.~\ref{fig:f_3panel}(b) and Fig.~\ref{fig:f_3panel}(c) also suggests that random sequences are always more efficient than alternate sequences. These basic observations are confirmed in Fig.~\ref{fig:f_quotiens_3panel}, where the ratios between fluctuation correction factors are displayed for all possible pairs (i.e. $F^{+-}/F^{++}$, $F^\mathrm{rnd}/F^{++}$, $F^{+-}/F^\mathrm{rnd}$). Panel (a) shows that, as was to be expected, $F^{++}$ is always more efficient than $F^{+-}$ (ratio always lower than 1). Similarly, panel (b) shows that $F^{++}$ is also more efficient than $F^\mathrm{rnd}$. Finally, panel (c) confirms that a random sequence is always more efficient than an alternate sequence, a logical consequence of the previous relation ($F^{+-}/F^{++} \leq 1$): if a random and an alternate sequence are compared, while they will display equally alternate pulses for some stretches of time, it will also be the case that for some stretches of time the random sequence will contain same sign pulses (i.e. $++$ or $--$), which will be more efficient than the corresponding alternate stretches. 

\begin{figure}[H]
    \centering
    \includegraphics[width=0.48\textwidth]{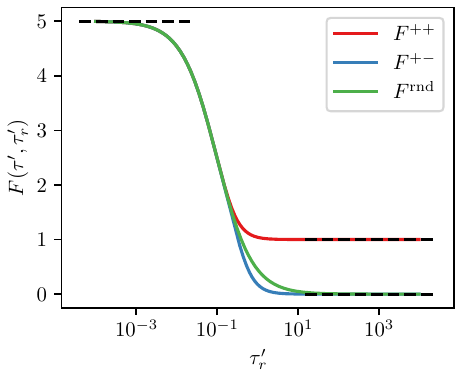}
    \caption{Fluctuation correction factors for exponential $++$ profiles (red), $+-$ (blue) and $\mathrm{rnd}$ (green) as a function of $\tau_r'$ for a fixed value of $\tau'=0.1$. Dashed black lines indicate the small and large value limits for each function (see second and fourth columns in Table \ref{tab:limits}).\label{fig:f_xB}}
\end{figure}

\begin{figure}[H]
    \centering
    \includegraphics[width=\textwidth]{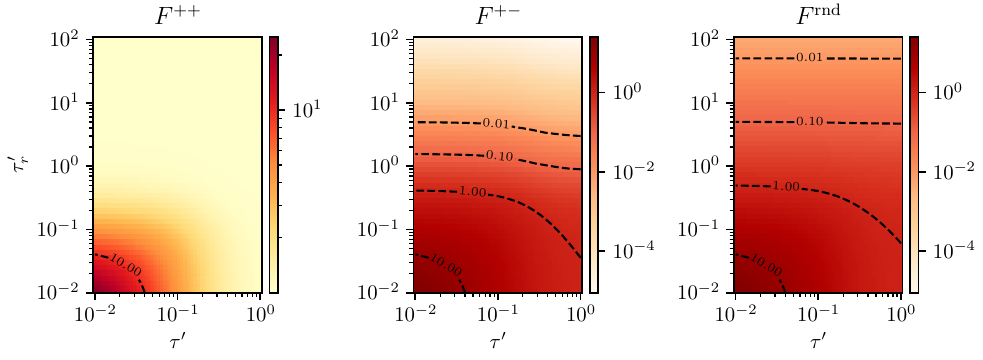}
    \caption{Contour maps of the fluctuation correction factors for exponential $++$ profiles (left), $+-$ (center) and $\mathrm{rnd}$ (right). Notice that the scale for the fluctuation correction factor differs between panel (a) and panels (b,c)), as the correction implied for $F^{++}$ is somewhat stronger for the region displayed. \label{fig:f_3panel}}
\end{figure}
\begin{figure}[H]
    \centering
    \includegraphics[width=\textwidth]{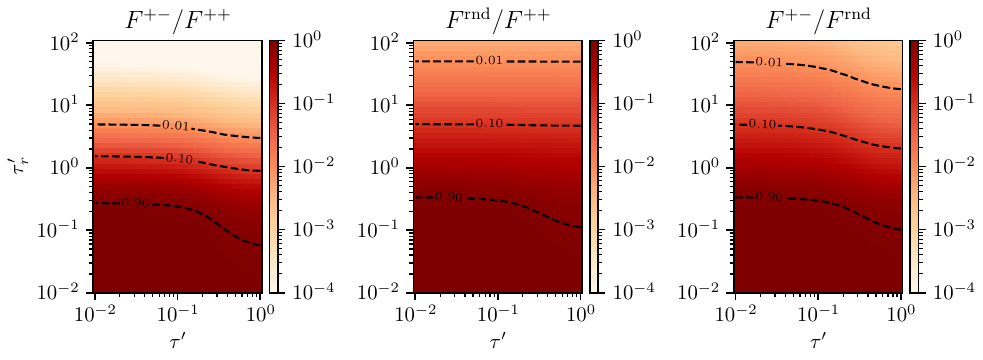}
    \caption{Contour maps of the ratios (see labels on the respective ordinate axes), of fluctuation correction factors for exponential $+-$ and $++$ profiles (left), $\mathrm{rnd}$ and $++$ (center), and $+-$ and $\mathrm{rnd}$ (right).\label{fig:f_quotiens_3panel}}
\end{figure}

Given the monotonic behaviour of the fluctuation correction factor, we only need to compute the limits of the closed-form formulas in order to complete the characterization of the whole two-dimensional parameter space.
These limits are collected in Table~\ref{tab:limits}, and reveal a more nuanced scenario than is apparent from the figures. It may be illustrative to start comparing the limits associated to varying $\tau_r'$ (for a fixed value of $\tau'$) with the corresponding results for harmonic pulses (for which $\tau_r'$ was the only variable of the fluctuation correction factor, see Fig.~\ref{fig:sin_absin_1} and related discussion). The full behaviour of the fluctuation correction factor for each sequence is depicted in Fig.~\ref{fig:f_xB}. The limit $\tau_r' \rightarrow \infty$ (rightmost column of Table~\ref{tab:limits}) corresponds to slow rotational diffusion compared with the period of the exponential kicks, and the results do not differ from what was obtained for harmonic signals (in any dimension): no diffusion enhancement for zero average velocity profiles ($F^{+-}=F^\mathrm{rnd} = 0$), and an enhancement indistinguishable from constant velocity for positive-defined pulses ($F^{++} = 1$). If $\tau_r'$ is decreased an inflexion point appears for $\tau_r' \approx 1$, as for the harmonic case. If it is further decreased ($\tau_r' \rightarrow 0$, central column in Table~\ref{tab:limits}), while there still are similarities with the harmonic case (all functions tend to the same value as the rotational parameter diminishes), the plateau is not a constant anymore (as it was for the harmonic case, $\pi^2/8$), instead it is now a function of pulse lifetime ($\tau'$). For the particular instance chosen in Fig. \ref{fig:f_xB} ($\tau' = 0.1$, i.e. pulse duration one tenth of time between exponentials $T$), this plateau corresponds to a correction factor of 5: \emph{any} exponential sequence would in this case result in a five-fold increase of the correction ($\mathrm{Pe}^2$), associated to constant velocity (equal to the average speed). 

\begin{table}[H]
\Large
\caption{Limit values of the fluctuation correction factor for exponential $++$ $+-$ and $\mathrm{rnd}$ profiles. \label{tab:limits}}
\begin{ruledtabular}
\begin{tabular}{ccccc}
 & \multicolumn{1}{p{3cm}}{\centering $\tau'\rightarrow 0$ \\ $(\tau << T)$} &  \multicolumn{1}{p{3cm}}{\centering $\tau_r'\rightarrow 0$ \\ $(\tau_r << T)$} & \multicolumn{1}{p{3cm}}{\centering $\tau'\rightarrow \infty$ \\ $(\tau >> T)$} & \multicolumn{1}{p{3cm}}{\centering $\tau_r'\rightarrow \infty$ \\ $(\tau_r >>T)$}\\\hline
$F^{++}$ & $\frac{1}{2\tau_r'}\frac{e^{1/\tau_r'}+1}{e^{1/\tau_r'}-1}$ & $\frac{1}{2\tau'}\frac{e^{1/\tau'}+1}{e^{1/\tau'}-1}$ & $1$ & $1$\\
$F^{+-}$ & $\frac{1}{2\tau_r'}\frac{e^{1/\tau_r'}-1}{e^{1/\tau_r'}+1}$ & $\frac{1}{2\tau'}\frac{e^{1/\tau'}+1}{e^{1/\tau'}-1}$ & $\frac{2\tau_r'\qty(1-e^{1/\tau_r'})+1+ e^{1/\tau_r'}}{1+e^{1/\tau_r'}}$ & $0$\\
$F^\mathrm{rnd}$ & $\frac{1}{2\tau_r'}$ & $\frac{1}{2\tau'}\frac{e^{1/\tau'}+1}{e^{1/\tau'}-1}$ & $\tau_r'\qty(e^{-1/\tau_r'}-1)+1$ & $0$\\
\end{tabular}
\end{ruledtabular}
\end{table}

\begin{figure}[H]
    \centering
    \includegraphics[width=0.48\textwidth]{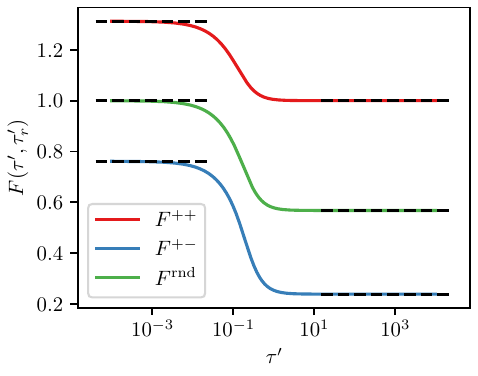}
    \caption{Fluctuation correction factors for exponential $++$ profiles (red), $+-$ (blue) and $\mathrm{rnd}$ (green) as a function of $\tau'$ for a fixed value of $\tau_r'=0.5$. Dashed black lines indicate the small and large value limits for each function.\label{fig:f_xA}}
\end{figure}

One could also ask for the behaviour that results from fixing $\tau_r'$, and allowing for the variation of $\tau'$ (i.e. a horizontal cut in Fig.~\ref{fig:f_3panel}, while the previous analysis corresponded to a vertical cut, for each panel). This possibility was not available for harmonic pulses as their ``lifetime'' ($\approx T/2$) was entirely fixed by the associated period (which would loosely correspond to a constant value $\tau' \sim 0.5 $). For exponential pulses the limit $\tau' \rightarrow \infty$ can be found in the second to rightmost column of Table~\ref{tab:limits}. There is no advantage (with respect to constant velocity) for same-sign sequences ($F^{++} =1$ irrespective of the rotational lifetime). In contrast, for zero average sequences there is always an advantage ($F^{+-}, F^\mathrm{rnd} \geq 0$), which depends on the rotational lifetime (as long as it is different from zero). We recall that for both these cases the corresponding average velocity is zero, and it is with respect to this case that the advantage exists. This advantage never surpasses though the diffusion enhancement obtained from the case at constant velocity (equal to the average speed), as a simple analysis of the analytic expressions in Table~\ref{tab:limits} also confirms). The corresponding behaviour is illustrated in Fig.~\ref{fig:f_xA}, with the zero average velocity cases corresponding to the alternate (blue) and random (green) sequences. When $\tau'$ is decreased, an inflexion point is found for $\tau'\approx1$ after which, upon further decrease of $\tau'$, the correction factor increases for all three cases, without coalescing (in contrast with the behaviour found in Fig.~\ref{fig:f_xB}). 

The previous discussion focused on horizontal/vertical cuts on Fig.~\ref{fig:f_3panel} and therefore the limits obtained depended on the particular value of the variable left fixed. This can be clearly seen from the (monotonic) functions in Table~\ref{tab:limits}, only for a handful of the cases the result is a constant (0,1). It is therefore of interest to obtain the margins of variation of each limiting function (i.e. compute the double limits).  

Before proceeding, it should be noted that these limits describe idealized scenarios. Their physical requirements should not be forgotten, considering that the associated effective diffusion is related to the product of the correction factor ($F$) and the square of the average P\'eclet number ($\avgPe$). In order to further illustrate this point with an example, let us consider the expression of $\avgPe$ for exponential profiles 
\begin{equation}
    \avgPe^2 = \frac{v_p^2 }{2 D_T D_R} (\tau')^2 \qty(1 - \exp^{-1/\tau'})^2 . \label{eq:mean-pe-2}
\end{equation}
We have for instance that a limit $ \tau' \rightarrow 0$ (with the reasonable assumption that $D_T,D_R$ are not modifiable), would imply that $v_p$ should be increased accordingly (in order to maintain $\avgPe$ constant). Of course there will be physical limits to the maximum value that can be achieved, depending on the workings of the system under study. Therefore the corresponding idealized limit will, generally, not be reachable. With this important consideration in mind, in the numerical examples to be discussed below, only modest changes from constant propulsion will be considered (i.e. $F \gtrsim 1$). 

With these provisos, the resulting limits are summarized in Table~\ref{tab:limits-dobles} (none of the double limits depends on the order taken), and can be grouped into just three distinct (idealized) regimes.

\begin{table}[H]
\Large
\caption{Double limit values of the fluctuation correction factor for exponential $++$ $+-$ and $\mathrm{rnd}$ profiles. \label{tab:limits-dobles}}
\begin{ruledtabular}
\begin{tabular}{ccccc}
&    \multicolumn{1}{p{4cm}}{\centering $\tau_r'\rightarrow \infty$ (any $\tau'$) }&  
  \multicolumn{1}{p{4cm}}{\centering   $\tau_r'\rightarrow 0, \tau'\rightarrow \infty$}
 & \multicolumn{1}{p{4cm}}{\centering  $\tau_r',\tau'\rightarrow 0$ }  \\\hline
$F^{++}$ & $1$ & $1$ & $\rightarrow \infty$ \\
$F^{+-}$ & $0$ & $1$ & $\rightarrow \infty$ \\
$F^\mathrm{rnd}$ & $0$ & $1$ & $\rightarrow \infty$ 
\end{tabular}
\end{ruledtabular}
\end{table}

\subsubsection{Frozen orientation regime}
The first column in Table~\ref{tab:limits-dobles} corresponds to the physical scenario where the orientation of the swimmer barely changes in between kicks. As it was the case for harmonic profiles, no enhancement is observed with respect to a constant velocity (equal to the average value), for none of the propulsion sequences. The same sign sequence ($++$) is equivalent to constant velocity ($F=1$), while the zero average sequences ($+-$ and random) have null enhancement ($F=0$) with respect to simple center of mass diffusion. The latter case ($+-$ and random) is aligned with the scallop theorem, according to which reversible sequences do not produce propulsion, if rotational fluctuations can be neglected. It is interesting to note that the present results indicate that this statement might need to be generalized to include random sequences as well, even though they are not time reversible. 

Finally, the damping effect of very slow rotational diffusion (with respect to actuation period, T) affects all three sequences, as the result is independent of the propulsion lifetime. The rightmost column in Table~\ref{tab:limits} was actually summarizing this situation, as no dependence on pulse lifetime ($\tau'$) was apparent either. In the opposite limit ($\tau_r' \rightarrow 0$), the behaviour will critically depend on the pulse's lifetime, and therefore two distinct limits will need to be considered.

\subsubsection{Orientationally randomized constant propulsion regime}
The second and last columns in Table~\ref{tab:limits-dobles} both correspond to fast rotational diffusion (with respect to the propulsion period), i.e. there is full orientational decorrelation in between kicks. In practical terms this implies that whether propulsion maintains its direction (with respect to the swimmer axis), or inverts it, will be of no consequence, all propulsion sequences will be equivalent. This is reflected in the fact that the corresponding boxes (for $++$, $+-$, $\mathrm{rnd}$ sequences), in each of the two last columns (different limits of the pulse lifetime $\tau'$), have equal values. For \emph{both} limits (large/small $\tau'$) the typical trajectory will be characterised a sequence of identical time intervals during which the swimmer is respectively propelled with constant/localized propulsion, with velocity direction randomization in between. As such, a description which applies to both limits might be that of ``run-and-tumble'' dynamics. One must be careful to note, though, that this terminology usually implies propulsion times which are not uniformly distributed, contrary to the present case (with a regular actuation). While a direct identification cannot thus be claimed at this stage, it seems clear that they constitute closely related instances of the same dynamical behaviour. 

In this section we focus on the second column, which corresponds to a large pulse lifetime with respect to the actuation period ($\tau' \rightarrow \infty $). The large pulse lifetime makes the kick indistinguishable from constant velocity. Consequently, no advantage results ($F=1$) with respect to constant velocity dynamics (the middle column is filled with unit values), and no further analysis is required. 

\subsubsection{Orientationally randomized propulsion bursts regime}
The last column in Table~\ref{tab:limits-dobles} displays what arguably constitutes the most interesting case, with a correction factor that diverges in all cases. This core result, a boundless enhancement in diffusion, contrasts with the results previously discussed for harmonic pulses, for which no remarkable enhancement was found with respect to constant propulsion. Such marked enhancement might have significant implications, as it suggests a possible strategy to overcome a well-known roadblock for constant propulsion velocity, namely that as the swimmer is downsized diffusion enhancement progressively disappears.\cite{Bechinger2016} This conclusion stems from analysis of Eq.~\ref{eq:d-vconst}, where the excess part ($v^2 \tau_r /2$, in 2-d) is inversely proportional to the rotational diffusion coefficient. Assuming a Debye-Einstein dependence for $D_R$ (which holds to a good approximation down to molecular scales), it follows that the excess diffusion coefficient will decrease as the cubic power of the swimmer radius. Consequently, diffusion enhancement from constant propulsion will become negligible for small enough swimmers (unless a large enough speed can be sustained). 

The present result, though, indicates that such decrease might be mitigated by using a train of exponential impulses instead of a constant velocity. The rationale at this point seems clear: generally, in order to avoid the damping effect of rotational diffusion, it is capital that swimmer displacements take place on time scales such that the swimmer has barely  changed its orientation. If rotational diffusion increases as well, this argument demands that the lifetime of the exponential kicks should decrease accordingly. As the formulas derived here show, the combined effect of this double limit ($\tau',\tau_r'\rightarrow 0$) results in an unbounded increase of the fluctuation correction factor ($F\rightarrow \infty$). 

It should be kept in mind though that the downside will be a corresponding high (but temporary) peak velocity ($v_p$), as previously discussed, in order to keep a constant $\avgPe$. A quantitative estimation seems appropriate at this point. We recall that we are comparing constant propulsion with exponential pulses, such that both have the same (average) P\'eclet number. If Eq.~\ref{eq:mean-pe-2} is thus equated to $\mathrm{Pe}^2$ (for constant propulsion), we obtain a simple relation between average and peak velocity (in the present $\tau' \rightarrow 0$ limit): $v \approx \tau' v_p$ . In addition, and according to the formulas displayed in the second column of Table \ref{tab:limits-dobles} (which are independent of sequence), if one seeks for instance a \emph{five-fold} increase of the effective diffusion coefficient ($F = 5$) with respect to constant propulsion, a value $\tau' = 0.1$ is required (for the limit case $\tau_r' \rightarrow 0$). Together with the previous relation linking peak and average speeds, we obtain $v_p\simeq 10 v$. That is, a five-fold enhancement would require a peak speed one order of magnitude larger than the corresponding constant speed, although the peak velocity just needs to be maintained for less than 10\% of the time, compared to constant velocity. As previously noted, there will be a physical feasibility limit that will depend on the system under study. Although it might seem a strong requirement, we emphasize that the present estimation corresponds to a substantial enhancement, a non-negligible advantage could still be obtained with more modest requirements as will be illustrated within the following section. 

In what concerns the dynamics, the (idealized) trajectory that may be expected in this regime is one that begins with a propulsion outburst which decays exponentially, followed by a period of inactivity ($\tau'\rightarrow 0$ implies $\tau << T$) during which damped diffusive dynamics ($D_\mathrm{eff} \approx D_T$) dominates, ending with a new propulsion outburst. It is capital that this second outburst departs in a \textit{random} direction with respect to the previous propulsion direction (as required by $\tau_r' \rightarrow 0$, i.e. small correlation time for rotation). 

\subsection{Numerical simulations}

\begin{figure}[H]
    \centering
    \includegraphics[width=\textwidth]{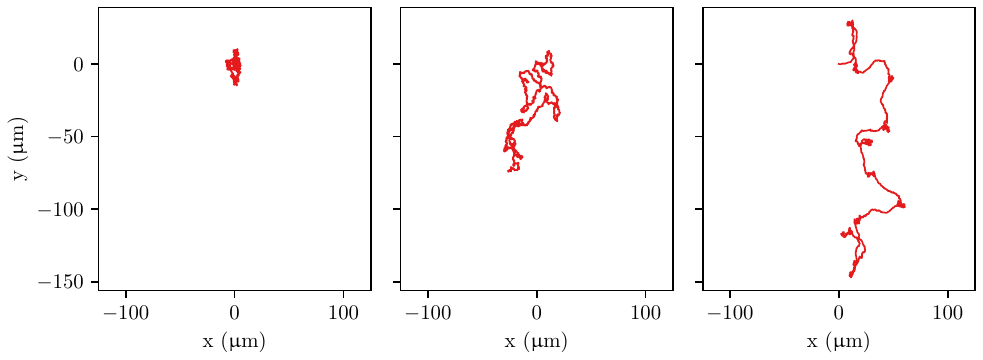}
    \caption{Single particle trajectories obtained \textit{via} stochastic simulations for free diffusion (left), constant velocity with a $\mathrm{Pe}\simeq3.72$ (center) and exponential $++$ with $\avgPe\simeq3.77$ and $F\simeq2$ (right). In all cases $D_T = 0.22\:\upmu \mathrm{m}^2/\mathrm{s}$, $D_R = 0.16 \:\mathrm{rad}^2/\mathrm{s}$ and a total simulation time of $400\:\mathrm{s}$. \label{fig:trajs}}
\end{figure}

In order to check these ideas and gain further insight into the characteristics of the regimes considered in the previous section, we provide here an account of selected simulation results {in 2-d} (the numerical techniques employed are to be found in SI). Although the discussion so far has been independent of scale, for the simulations we have particularized to a typical micrometric swimmer of radius $R=1\:\upmu\mathrm{m}$ immersed in water (with viscosity $\eta=10^{-3}\:\mathrm{N}\:\mathrm{s}/\mathrm{m}^2$ at $T=300\:\mathrm{K}$). Assuming the validity of the Einstein relations for translational and rotational diffusivity, this results in $ D_T \simeq 0.22\:\upmu\mathrm{m}^2/s $, and $D_R \simeq 0.16\:\mathrm{rad}^2/s $.
The left panel in Figure \ref{fig:trajs} depicts a sample trajectory with these parameters, i.e. with no propulsion, and is thus representative of the displacements characteristic of ``free'' micrometric spherical particles (over a 400 s time lapse). 

\begin{figure}[H]
    \centering
    \includegraphics[width=0.48\textwidth]{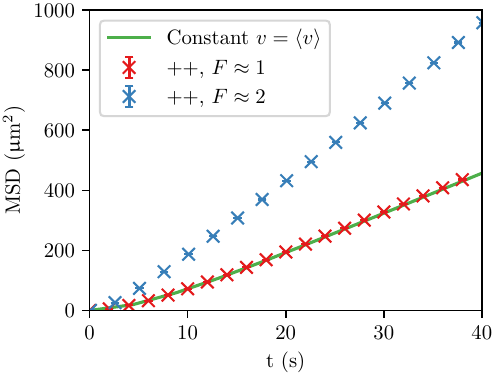}
    \caption{Mean Square Displacement for exponential $++$ profiles with parameters such that $F\simeq1$ (red crosses) and $F\simeq2$ (blue crosses), see Table.~\ref{tab:numerical}, obtained \textit{via} stochastic simulations. Theoretical results for constant velocity for the corresponding $\avgPe$ is also shown (green line). In all cases $D_T = 0.22\:\upmu \mathrm{m}^2/\mathrm{s}$, $D_R = 0.16 \:\mathrm{rad}^2/\mathrm{s}$ and a total simulation time of $40\:\mathrm{s}$ for constant and $F\simeq1$ and for $400\:\mathrm{s}$ $F\simeq2$ (only $40\:\mathrm{s}$ shown). For the exponential sequences  $v_p=10\:\upmu\mathrm{m/s}$, which corresponds to $\avgPe\simeq3.77$. \label{fig:msd_exp}}
\end{figure}

The middle panel in Fig.~\ref{fig:trajs} displays a sample trajectory after a \emph{constant} propulsion velocity has been added. A value $v=1\:\upmu\mathrm{m}/\mathrm{s}$ has been chosen, as it corresponds to the lower bound of the range of velocities reported for the first self-thermophoretic Janus particle described experimentally.~\cite{Jiang2010} It is also a small value if compared with typical artificial swimmer speeds, which usually fall in the tens of $\upmu\mathrm{m}/\mathrm{s}$ (as can be gathered for instance from the extensive list in Table I of Ref. \onlinecite{Bechinger2016}). Still, the effects on the dynamics of the micrometric particle are easily discerned in Fig.~\ref{fig:trajs}, with a substantially larger spread of the trajectory when compared with the free particle (leftmost panel). The corresponding (theoretical) MSD is displayed in Fig.~\ref{fig:msd_exp} (green line). The associated P\'eclet number is $\mathrm{Pe}\simeq3.77$, so the effective diffusion coefficient that results is $D_\mathrm{eff} = D_T \qty(1 + \mathrm{Pe}^2)  \simeq 15 D_T$. In short, even with a rather small value of $v$, the effective diffusion coefficient turns out to be a factor $\simeq 15$ times larger than that characteristic of free diffusion ($D_T$).

We now turn to the impact of using a time varying propulsion velocity characterized by exponential kicks. The rightmost panel in Fig.~\ref{fig:trajs} displays a sample trajectory for an exponential $++$ sequence. It is equally characterized by $\avgPe\simeq3.77$, and a modest correction factor of value $F\simeq2$. This correction factor results from picking $\tau=4\:\mathrm{s}$ and $T=40\:\mathrm{s}$, i.e. the time in between pulses is ten times larger than the duration of the pulse itself. In terms of the corresponding scaled characteristic times, we have that for this choice $\tau' = 0.1$ and $\tau_r \simeq 0.16$, i.e. similarly small values representative of the idealized scenario discussed in the previous section. It might have been noticed that the scaled pulse lifetime ($\tau' = 0.1$) is identical to the numerical example discussed in the previous section for order of magnitude estimations, and yet, the corresponding correction factor is now smaller ($F(0.1,0.16) \simeq 2$ instead of the previous $F(0.1,0)\simeq 5$). This difference is to be attributed to the fact that for the present, more realistic, case the associated rotational relaxation time is different from zero (as is evident from comparing the arguments of the correction function in the previous parenthetical remark). Turning to the effective diffusion coefficient that results for this set of parameters, according to Eq.~\ref{eq:d-pe} we have $D_\mathrm{eff} = D_T \qty[1+ \avgPe^2 F\qty(0.1,0.16)]  \simeq 29.5 D_T$, i.e. the factor for constant velocity ($\simeq 15$) has been doubled. This effect is evident in Fig. \ref{fig:msd_exp}, where the corresponding MSD (blue crosses) is compared to constant velocity (green line). 

Direct inspection of the corresponding trajectory (rightmost panel in Fig.~\ref{fig:trajs}) shows, in line with what had been anticipated in the previous section, a sequence of long excursions punctuated by localized tangled knots (during which rotational diffusion randomizes the swimmer's orientation). Therefore, besides the doubling of diffusion, the impact of the correction factor on the trajectory characteristics is remarkable, particularly considering that a rather modest pulse to period ratio has been considered ($\tau'\simeq\tau_r'\simeq 0.1$). In particular, the present estimation suggests that for the Janus particle in Ref.~\onlinecite{Jiang2010}, where constant illumination was used, a pulsed sequence might prove to be advantageous, assuming exponentially decaying pulses. The present, rough, estimation indicates that a sequence of pulses (lifetime 4 s) with a period of 40 s, might result in a doubling of the diffusion coefficient, and a distinct qualitative change of the trajectories. While the required pulses should be capable of producing a peak speed of $10 \; \upmu\mathrm{m}/\mathrm{s}$, we note that in Ref.~\onlinecite{Jiang2010} the ability to produce sustained speeds up to $6 \:\upmu\mathrm{m}/\mathrm{s}$ was reported.

Table~\ref{tab:numerical} contains the (numerical and theoretical) effective diffusion coefficients for the case just discussed (within the column with heading $\tau = 4\:\mathrm{s}$, $T = 40\:\mathrm{s}$), and for the constant velocity propulsion as well (first row). In addition, the results for several other cases are also displayed, corresponding to different sequences ($+-$ and random) and sets of parameters ($\tau = 0.1\:\mathrm{s}$, $T = 1\:\mathrm{s}$, for which $F\simeq1$). The middle column illustrates how for the latter set of parameters no difference is found between a $++$ sequence (first box) and constant speed (last box), both with an effective diffusion coefficient $D_\mathrm{eff}\simeq 3.35$. In addition, and still for this set of parameters ($\tau = 0.1\:\mathrm{s}$, $T = 1\:\mathrm{s}$), the $+-$ sequence barely differs from free diffusion ($D_\mathrm{eff} \simeq D_T$), while the random sequence is characterized by double this value, consistently with its superior efficiency. The last column in Table~\ref{tab:numerical} shows that for $\tau'\simeq\tau_r'\simeq 0.1$ all sequences are essentially equivalent. Quantitative agreement between theory and numerical results is excellent in all cases, including the case of random signed pulses, for which only a closed-formed solution is available. We note that the MSD for the $++$ sequence with $F\simeq1$ is also displayed in Fig.~\ref{fig:msd_exp} (red crosses, coincident with constant propulsion).  Finally, in order to provide a more complete picture of the different regimes, trajectories for other values of $\avgPe$ and $F$ can be found in SI.

\begin{table}[H]
\caption{\label{tab:numerical} 
Effective diffusion coefficients obtained \textit{via} stochastic simulations {in 2-d} with $D_T = 0.22\:\upmu \mathrm{m}^2/\mathrm{s}$ and $D_R = 0.16 \:\mathrm{rad}^2/\mathrm{s}$ for the two sets of parameters studied.}
\begin{ruledtabular}
\begin{tabular}{l|ll|l}
 & & \parbox[b]{3cm}{$\tau = 0.1\:\mathrm{s}$\\$T = 1\:\mathrm{s}$} & \parbox[b]{2.5cm}{$\tau = 4\:\mathrm{s}$\\$T = 40\:\mathrm{s}$}\\\colrule
 \multirow{2}{*}{ct}& $D_\mathrm{sim}$               & $3.34\pm0.02$ & $3.34\pm0.02$\\
 &$D_\mathrm{theo}$ & $3.35$       & $3.35$\\\colrule
\multirow{2}{*}{++}& $D_\mathrm{sim}$               & $3.33\pm0.02$ & $6.38\pm0.02$\\
 &$D_\mathrm{theo}$ & $3.35$ ($F\simeq1$)        & $6.37$ ($F\simeq1.97$)\\\colrule
\multirow{2}{*}{+-}& $D_\mathrm{sim}$               & $0.236\pm0.001$ & $6.27\pm0.01$\\
 &$D_\mathrm{theo}$ & $0.236$ ($F\simeq5\cdot10^{-3}$)        & $6.26$ ($F\simeq1.93$) \\\colrule
\multirow{2}{*}{rnd}& $D_\mathrm{sim}$               & $0.467\pm0.002$ & $6.34\pm0.02$\\
 &$D_\mathrm{theo}$ & $0.466$ ($F\simeq0.08$)       & $6.32$ ($F\simeq1.95$)\\
\end{tabular}
\end{ruledtabular}
\end{table}

\section{Concluding Remarks}
The goal of this work has been the assessment of optimal propulsion strategies for diffusion enhancement, building upon the basic finding that a periodic propulsion velocity is in all cases more efficient than the corresponding average velocity propulsion. This enhancement has been shown to critically depend on the propulsion motif that is repeated. While a harmonic profile results in a marginal advantage, an exponential profile turns out to allow for controllable speed-ups. A complete characterization of the parameter landscape associated to a train of exponential impulses shows that the optimum, idealized, strategy consists of a sequence of delta like propulsion kicks. These should be sufficiently separated in time so that orientational correlation is lost in between successive impulses. The consideration that a similar orientational decorrelation does not provide any remarkable advantage for harmonic sequences, leads to the realization that the ability to control the ratio between pulse lifetime and actuation period is essential. The potentially large diffusion enhancements that result suggest a possible strategy to dampen the known deleterious effect of rotational diffusion as the swimmer is downsized. The small molecule limit, in which we are currently exploring different propulsion mechanisms controlled by external radiation,\cite{Calero2020,Jurado2023} might constitute an scenario where such periodic excitation protocol might play a crucial role. Finally, a delta like limit can be reached with different functional forms, with the present choice (exponential profile) being dictated by recent findings for a variety of swimmer types and length scales. It is to be expected that equivalent (in the delta limit) profiles, might result in comparable advantages for realistic parameter sets.

\begin{acknowledgments}
AJR acknowledges the financial support from Departament de Recerca i Universitats de la Generalitat de Catalunya. CC acknowledges support from the Spanish grant PID2021-124297NB-C31 funded by MCIN/AEI/10.13039/501100011033 and “ERDF A way of making Europe”, and from the Spanish grant CNS2022-135395 funded by MCIN/AEI/10.13039/ 501100011033 and by the European Union NextGenerationEU/PRTR. AJR and RR acknowledge support from grant PID2021-124297NB-C32, funded by MCIN/AEI/10.13039/501100011033 and FEDER. The authors thankfully acknowledge RES resources provided in machine Xula to FI-2023-2-0011.
\end{acknowledgments}

\bibliography{biblio}

\begin{widetext}
\section{Appendix: Closed-form effective diffusion coefficients}
Here we show how a closed-form solution can be obtained for a propulsion velocity which consists of a sequence of exponential pulses (regular in time and signed). The pulse for a generic interval ($\delta +i T\leq t <\delta+ (i+1) T$) is defined as

\begin{equation}
v(t) = a_i v_p e^{-\beta\qty(t-\delta -i T)}  ,
\end{equation}
where $\beta \equiv 1/\tau$ and $a_i= \pm 1 $ (with the specific sign depending on the sequence of interest). In addition, a time shift ($\delta$) has been introduced in order to allow for a final average over time-shifted origins (a procedure that will depend on the specific sequence of signs for the successive exponentials). Figure \ref{fig:int} displays a time-shifted sequence of positively defined kicks ($a_i=1$).

\begin{figure}[H]
    \centering
    \includegraphics[width=\textwidth]{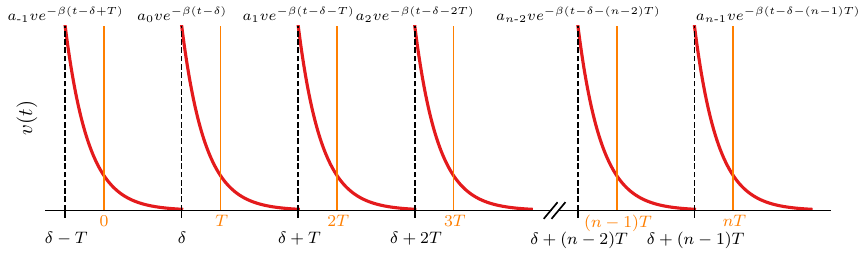}
    \caption{Sequence of positively defined exponential pulses shifted by a time $\delta$. \label{fig:int}}
\end{figure}

The quantity we will focus on is the derivative of the MSD (Eq. \ref{eq:dbasic}) and, in particular, its non-trivial part 
\begin{equation}\label{eq:integral}
    \left\langle  v\qty(t)e^{-t/\tau_r} \int_0^t \dd{\uptau} v\qty(\uptau)e^{\uptau/\tau_r} \right\rangle_{v} \rm.
\end{equation}

For a time $t \in \qty[\delta + (n-1) T,\delta + nT]$, the integral within this average is given by
\begin{align}
   \int_0^t v(\uptau) e^{\uptau/\tau_r} \dd{\uptau} = v_p e^{\beta \delta} \Big\{ \Big.  a_1 e^{-\beta T}I(0,\delta) 
    & + a_0 e^{-\beta T}I(\delta,\delta + T)+\\ 
    & + a_1 e^{\beta T}I(\delta+T,\delta+ 2T) + \cdots 
       + a_{n-1} e^{i\beta T}I(\delta + (n-1) T,t) \Big. \Big\} ,
\end{align}
as can be easily obtained with the help of Fig. \ref{fig:int}, and where the following integral has been defined
\begin{equation}
    I(a,b) \equiv \int_a^b e^{\qty(\tau_r^{-1}-\beta) t} \dd{t} .
\end{equation}

If we now include the prefactor of the integral in Eq. \ref{eq:integral}, and again for $t \in \qty[\delta + (n-1) T,\delta + nT]$,
\begin{align}
    v\qty(t)e^{-t/\tau_r} \int_0^t \dd{\uptau} v\qty(\uptau)e^{\uptau/\tau_r}& =  v_p^2 e^{2 \beta \delta+ \beta (n-2)T - \qty(\beta + \tau_r^{-1}) t}\Big\{ \Big.  a_{n-1}a_{-1}I(0,\delta) +  \\
    &  + a_{n-1}a_0 e^{\beta T} I(\delta,\delta+ T)+a_{n-1}a_1 e^{2\beta T} I(\delta+T,\delta+ 2T)+ \cdots \\
      \cdots & + a_{n-1}a_{n-2} e^{(n-1)\beta T} I\qty[\delta+(n-2)T,\delta+(n-1) T]+ e^{n\beta T} I\qty[\delta+(n-1)T,t] \Big. \Big\} ,
\end{align}

which after writing the explicit form of the integrals, and grouping terms, results in
\begin{align}
    v\qty(t)e^{-t/\tau_r} \int_0^t \dd{\uptau} v\qty(\uptau)e^{\uptau/\tau_r} & =  \frac{v_p^2 e^{2 \beta \delta+ \beta (n-2)T - \qty(\beta + \tau_r^{-1}) t}}{\tau_r^{-1}-\beta} \Big\{ \Big. e^{\qty(\tau_r^{-1}-\beta)\delta}a_{n-1} \qty[ a_{-1}+ a_0 e^{T/\tau_r}+ \cdots+ a_{n-2} e^{(n-1)T/\tau_r}] + \\     
    & - e^{\qty(\tau_r^{-1}-\beta)\delta}a_{n-1} e^{\beta T} \qty[ a_0 +a_1 e^{T/\tau_r}+ \cdots+ a_{n-2} e^{(n-2)T/\tau_r}] +\\
    & - a_{-1}a_{n-1}+ e^{n\beta T+ (\tau_r^{-1}-\beta)t} -e^{\qty(\tau_r^{-1}-\beta)\delta}e^{(n-1)T/\tau_r+\beta T}
    \Big. \Big\} .
\end{align}

Finally, the average over origins $<>_v$ amounts to performing an integral $\frac{1}{T}\int_0^T d\delta$ for all terms, from which 
\begin{align}
\left\langle v\qty(t)e^{-t/\tau_r} \int_0^t \dd{\uptau} v\qty(\uptau)e^{\uptau/\tau_r}\right\rangle_v & =   \frac{v_p^2 e^{\beta (n-2)T - \qty(\beta + \tau_r^{-1}) t}}{\tau_r^{-1}-\beta} \Bigg\{ \Bigg. \frac{e^{\qty(\tau_r^{-1}+\beta)T}-1}{\qty(\tau_r^{-1}+\beta)T} \Big\{ \Big. a_{n-1}a_{-1} + 
a_{n-1}\qty(e^{t/\tau_r}-e^{\beta T}) \nonumber \\ & \qty[ a_0 +a_1 e^{T/\tau_r} \cdots+ a_{n-2} e^{(n-2)T/\tau_r}] - e^{\beta T + \qty(n-1)T/\tau_r} \Big. \Big\} + \nonumber \\ & \qty[ -a_{-1}a_{n-1} + e^{n\beta T + \qty(\tau_r^{-1}-\beta)t}]\frac{e^{2 \beta T}-1}{2 \beta T} 
     \Bigg. \Bigg\}  .  \label{eq:average-excess}
 \end{align}
We now particularize to the three cases of interest.

\subsection{Sequence of pulses with same sign }
For the $++$ sequence we just need to set $a_i=1$ in the previous expression (Eq. \ref{eq:average-excess}), from which
\begin{align}
\left\langle v\qty(t)e^{-t/\tau_r} \int_0^t \dd{\uptau} v\qty(\uptau)e^{\uptau/\tau_r}\right\rangle_v & =   \frac{v_p^2 e^{\beta (n-2)T - \qty(\beta + \tau_r^{-1}) t}}{\tau_r^{-1}-\beta} \Bigg\{ \Bigg. \frac{e^{\qty(\tau_r^{-1}+\beta)T}-1}{\qty(\tau_r^{-1}+\beta)T} \Big\{ \Big. 1 + 
\qty(e^{t/\tau_r}-e^{\beta T}) \frac{1-e^{\qty(n-1)T/\tau_r}}{1-e^{T/\tau_r}} \\ & - e^{\beta T + \qty(n-1)T/\tau_r} \Big. \Big\} + \qty[ -1 + e^{n\beta T + \qty(\tau_r^{-1}-\beta)t}]\frac{e^{2 \beta T}-1}{2 \beta T}
    \Bigg. \Bigg\}    .
\end{align}
It should be recalled at this point that this formula corresponds to the interval $t \in \qty[\delta + (n-1) T,\delta + nT]$. As previously noted, the average over origins corresponds to $\delta$ spanning a full period ($\delta \in [0,T]$). Consequently, upon variation of $\delta$, the time $t= nT$ samples all the possible values of the time-shifted exponential profile. Therefore, it is only required to particularize the formula above to $t= nT$ and perform the limit $n\rightarrow \infty$, in order to obtain the averaged excess diffusion coefficient ($D_\mathrm{ex}$), free from transients.  The result of this limiting procedure is 
\begin{equation}
\left\langle v\qty(t)e^{-t/\tau_r} \int_0^t \dd{\uptau} v\qty(\uptau)e^{\uptau/\tau_r}\right\rangle_v \xrightarrow{t= nT \rightarrow \infty } \frac{v_p^2}{\qty(\tau_r^{-1}-\beta)} \qty{ \frac{\qty[e^{-2\beta T}-e^{\qty(\tau_r^{-1}-\beta) T}]\qty(1-e^{\beta T})}{T \qty(\tau_r^{-1}+\beta)\qty(1-e^{T/\tau_r})}+\frac{1-e^{-2T/\tau_r}}{2\beta} } ,
\end{equation}
from which we can extract the excess contribution to the diffusion coefficient
\begin{equation}
    D^{++}_\mathrm{ex} = D_T {\frac{2}{d(d-1)}}\langle Pe \rangle^2
    \qty{\frac{\qty(T/\tau)^2}{\qty(1-\frac{\tau_R}{\tau})\qty(1-e^{-T/\tau})^2}
    \qty[\frac{\qty(e^{-2T/\tau}-e^{T/\tau_R-T/\tau})\qty(1-e^{T/\tau})}{\qty(\frac{T}{\tau_R}+\frac{T}{\tau})\qty(1-e^{T/\tau_R})}+\frac{1-e^{-2T/\tau}}{2\frac{T}{\tau}}]} \rm .
\end{equation}

The term in braces can be readily identified with one of the instances of the correction factor ($F(\tau',\tau_r')$) in Eqs. (\ref{eq:fclosed},\ref{eq:gfactor}), which has been the subject of a detailed analysis in Sec. \ref{sec:param_maps}. The correctness of this expression has been checked by comparison with the corresponding analytic (Fourier) formula (Eqs. \ref{eq:fpp}, see SI), and with sample numerical simulations (see Table \ref{tab:numerical}).

\subsection{Sequence of pulses with alternate sign }
We start with a slightly more general formulation (that would include the previous, same sign, case), by setting: $a_i \equiv A$ (for $i=\cdots,0,2,\cdots$), and $a_i \equiv B$ (for $i=\cdots,-1,1,3,\cdots$). The previous case corresponds to $A=B=+1$, while for the present case $A=-B=+1$. When substituted into Eq. \ref{eq:average-excess} we obtain

\begin{align}
&\left\langle v\qty(t)e^{-t/\tau_r} \int_0^t \dd{\uptau} v\qty(\uptau)e^{\uptau/\tau_r}\right\rangle_v  =   \frac{v_p^2 e^{\beta (n-2)T - \qty(\beta + \tau_r^{-1}) t}}{\tau_r^{-1}-\beta} \Bigg\{ \Bigg. \frac{e^{\qty(\tau_r^{-1}+\beta)T}-1}{\qty(\tau_r^{-1}+\beta)T} \Big\{ \Big. 1 + \\ & 
\qty(e^{t/\tau_r}-e^{\beta T}) \qty[ AB + e^{T/\tau_r} \cdots+ AB e^{(n-2)T/\tau_r}] - e^{\beta T + \qty(n-1)T/\tau_r} \Big. \Big\} + \qty[ -1 + e^{n\beta T + \qty(\tau_r^{-1}-\beta)t}]\frac{e^{2 \beta T}-1}{2 \beta T}
    \Bigg. \Bigg\}   , \label{eq:alternate}
\end{align}
which reveals an invariance with respect to the exchange $A \leftrightarrow B$, i.e. the result is the same whether the sequence starts with a positive pulse (followed by a negative one), or its inverse. We conclude that an average over $\delta \in [0,T]$ is independent of whether the first pulse is positive or negative (as long as they are subsequently alternate in sign). However, this is just a part of the average that we need to compute here.

If we consider for instance the $(-1,1,-1,\cdots)$ sequence, a full period consists of a negative amplitude exponential followed by a positive amplitude exponential. Consequently, we should allow for $\delta \in [0,2T]$ when performing the average (where we recall that $T$ denotes the duration of a single exponential). As we have seen, though, Eq. \ref{eq:alternate} is only valid for an average over $\delta \in [0,T]$. It is still required to allow for $\delta \in [T,2T]$, but this would be equivalent to shifting the whole sequence by a time $T$, i.e. turning the sequence into a $(+1,-1,\cdots)$ sequence, and averaging for $\delta \in [0,T]$. Considering the invariance previously found, this second part of the average will produce the same result. Therefore, the expression found in Eq. \ref{eq:alternate} is actually sufficient and no further averaging is required. 

Finally, concerning the upper time limit for the interval $t= n\Delta$, $n$ should be even, that is $n = 2 k$ in order to ensure that it corresponds to an integer number of repetitions of the basic $+-$ sequence. With these provisos we obtain
\begin{align}
&\left\langle v\qty(t)e^{-t/\tau_r} \int_0^t \dd{\uptau} v\qty(\uptau)e^{\uptau/\tau_r}\right\rangle_v  =   \frac{v_p^2 e^{-2 \beta T- 2 k \tau_r^{-1} T }}{\tau_r^{-1}-\beta} \Bigg\{ \Bigg. \frac{e^{\qty(\tau_r^{-1}+\beta)T}-1}{\qty(\tau_r^{-1}+\beta)T} \Big\{ \Big. 1 + \\ & 
\qty(e^{t/\tau_r}-e^{\beta T}) \qty[- \frac{1+e^{2 k T/\tau_r}}{1+e^{T/\tau_r}}] - e^{\beta T + \qty(2k-1)T/\tau_r} \Big. \Big\} + \qty[ -1 + e^{2k T/\tau_r}]\frac{e^{2 \beta T}-1}{2 \beta T}
    \Bigg. \Bigg\}  .
\end{align}

Finally, taking the limit $k \rightarrow \infty$, we get
\begin{equation}
    D^{+-}_\mathrm{ex} = D_T {\frac{2}{d(d-1)}}\langle Pe \rangle^2
    \qty{\frac{\qty(T/\tau)^2}{\qty(1-\frac{\tau_R}{\tau})\qty(1-e^{-T/\tau})^2}
    \qty[\frac{\qty(e^{-2T/\tau}-e^{T/\tau_R-T/\tau})\qty(1+e^{T/\tau})}{\qty(\frac{T}{\tau_R}+\frac{T}{\tau})\qty(1+e^{T/\tau_R})}+\frac{1-e^{-2T/\tau}}{2\frac{T}{\tau}}]} \rm .
\end{equation}
The term in braces can be readily identified with one of the instances of the correction factor ($F(\tau',\tau_r')$) in Eqs. (\ref{eq:fclosed},\ref{eq:gfactor}). This expression has also been checked against the analytic (Fourier) formula (Eq. \ref{eq:fpm}, see SI), and against numerical simulations as well (Table \ref{tab:numerical}).

\subsection{Sequence of pulses with random signs}
In this case we are not just considering a single sequence (with a variable time shift $\delta$), but all possible sequences with random amplitude signs (and with $\delta$ time shifts as well). That is, and additional average over all possible strings $\{a_i=\pm 1\}$ will be required. This can be easily performed by noting that in Eq. \ref{eq:average-excess} only bilinear products of amplitudes appear ($a_i\cdot a_j$), and that for random strings the following simple relations hold: $a_i^2 = 1$ (valid with/without averaging)  and $<a_i a_j> = 0$ (for $i\neq j$). When they are introduced into Eq. \ref{eq:average-excess} (i.e. when the average over strings is performed), we obtain for $t= nT$ (as for the same sign case)
\begin{equation}
    \left\langle v\qty(t)e^{-t/\tau_r} \int_0^t \dd{\uptau} v\qty(\uptau)e^{\uptau/\tau_r}\right\rangle_v = \frac{v_p^2}{\qty(\tau_r^{-1}-\beta)} \qty{ \frac{\qty[e^{-2\beta T}-e^{\qty(\tau_r^{-1}-\beta) T}]\qty(1-e^{\beta T})}{\qty(\tau_r^{-1}+\beta)\qty(1-e^{T/\tau_r})}+\frac{1-e^{-2T/\tau_r}}{2\beta} } ,
\end{equation}
i.e. with this extra averaging the expression is simplified to the point that no additional long time limit is required, as no time dependence remains. This is a remarkable property, implying that the diffusion coefficient can be obtained from very short time simulations: it is the value at $t=nT$ (for any value of $n$). In other words, with a simulation of length $T$ ($n=1$, i.e. just one pulse)  one can infer the (long time) diffusion coefficient. Supplementary Information contains a plot where this property is verified for a particular case. 

Returning to the closed-form expression, the excess part can be readily identified as
\begin{equation}
    D^{rnd}_\mathrm{ex} = D_T {\frac{2}{d(d-1)}}\langle Pe \rangle^2
    \qty{\frac{\qty(T/\tau)^2}{\qty(1-\frac{\tau_R}{\tau})\qty(1-e^{-T/\tau})^2}
    \qty[\frac{\qty(e^{-2T/\tau}-e^{T/\tau_R-T/\tau})}{\qty(\frac{T}{\tau_R}+\frac{T}{\tau})}+\frac{1-e^{-2T/\tau}}{2\frac{T}{\tau}}]} \rm ,
\end{equation}
which again is an instance of Eqs. (\ref{eq:fclosed},\ref{eq:gfactor}). As already stated, since a corresponding Fourier expression is not available in this case, only checks against numerical simulations are possible (see Table \ref{tab:numerical}).

\end{widetext}

\end{document}